# The ArduSiPM a compact trasportable Software/Hardware Data Acquisition system for SiPM detector.

Valerio Bocci , *Member, IEEE*, Giacomo Chiodi, Francesco Iacoangeli, Massimo Nuccetelli, Luigi Recchia

*Abstract*–The acquisition of a single Silicon Photomultiplier require multiple and expensive electronics modules as : preamplifier, discriminator, bias voltage power supply, temperature monitor, Scalers, Analog to Digital Converter and Time to Digital Converter . The developed ArduSiPM is a compact cost effective and easily replicable Hardware software module for SiPM detector readout.  The ArduSiPM uses an Arduino DUE (an open Software/Hardware board based on an ARM® Cortex™-M3 microcontroller) as processor board and a piggyback custom designed board (Shield), these are controlled by custom developed software and interface. The Shield contains different electronics features both to monitor, to set and to acquire the SiPM signals using the microcontroller board. The shield embed a controlled bias voltage power supply,  a fast voltage preamplifier, a programmable fast discriminator to generate over threshold digital pulse , a peak hold to measure the pulse height, a temperature monitor system, a scaler to monitor over threshold rate and arrival time of the pulses. A TCP/IP Wi-Fi or Ethernet connection is used both to control and to acquire the ArduSiPM remotely with either a PC software or a Tablet APP (depending on the application).

## I. INTRODUCTION

THE SiPM (Silicon Photon Multiplier)[6], also known as MPPC (multi pixel photon counter), is a solid photon counting device with single photon detection skill. The SiPM/MPPC is made up of an avalanche photodiode (APD) array on common Si substrate. Each single element working in Geiger mode. When a photon hits the silicon it produces a photoelectron through the photoelectric effect in one of the microcells. The Geiger regime create a large breakdown current: in this way each cell functions like a binary system. Thus, the summed current output of all the microcells is proportional to the number of fired cells, that means proportional to the number of detected photons. The signal generated from SiPM is in the order of one tenth of mV per photon. Respect to the classical Photo Multiplier, the SiPM is more compact, more robust and does not necessitate of in the KV order bias voltage that is difficult to manage in compact design. Indeed, the bias voltage of SiPM/MPPC changes from 30 V to less than 100 V depending on the series of device. The purpose of the ArduSiPM is to produce a portable, wireless, battery powered and compact system for use with different scintillating materials for particle's detection. The open hardware and software Arduino Due platform is a microcontroller board based on the Atmel SAM3X8E ARM Cortex-M3 CPU. The processor is a 32-bit ARM single core microcontroller. It has digital input/output pins, ADC analog inputs,  UARTs (hardware serial ports), a 84 MHz clock, DAC (digital to analog) output, eight 32 bits fast counters and controller lines for different serial bus (like I2C,SPI ). The Arduino Due system provides an open software to control the main operations but also a full access to the Atmel SAM3X8E control registers. We use this feature to obtain the best performance of the RISC processor. A TCP/IP stack for internet protocol communication is available but we choose to use a small WI-FI module with another ARM processor dedicated to data communication so as to avoid an overcharge of the main processor. The Arduino Due development board gives the access of all processor functionality via external connectors, which can be used to install piggyback board called Shield. The use of plugged shield gives the possibility to create an easy to handle by hand box. The ArduSiPM Shield we develop contains all the readout and control electronics to manage a SiPM, such as the power supply, or the temperature controller.

The purpose of the ArduSiPM system is both to generate a fast trigger output and monitor on board the main parameters of the signal such as arrival time, signal amplitude, number of counters in a specific time window, arrival time stamp of any over threshold signal. All controls setting and monitor data are available using a Wi-FI TCP/IP connection. We integrate in ArduSiPM a fast voltage amplifier to amplify and replicate the analog signal for the detailed study with scope. A fast discriminator with programmable threshold provide a trigger signal that can be used as output and, internally, as rate monitor and time stamp. The circuit contains a fast peak Hold ADC to measure pulse height a fast counter for rate measurements of over threshold events. The temperature monitor and the bias voltage are programmable and can be monitored for temperature compensation**.**

## II. THE ARDUSIPM ARCHITECTURE

The ArduSiPM system is constituted of three system : an open software/hardware Arduino DUE Board, a custom designed Arduino Shield, a Wifi module , a specific software to control all the function of the shield a client software. The Arduino DUE is an open hardware and software development board based on the Atmel SAM3X8E ARM Cortex-M3 CPU. The Arduino DUE schematics and the development software is available over the internet [1].

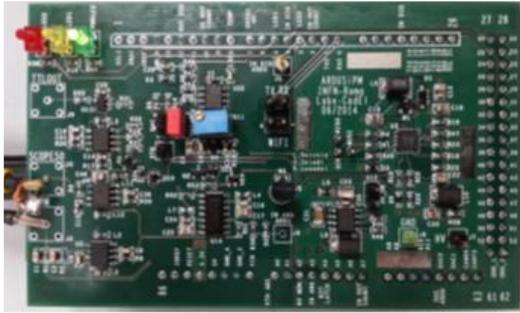

Fig. 1 ArduSiPM Shield

The Board can be connected with the external world via a series of connectors presents in the edge. A piggy board with the same form factor pluggable in Arduino is called Shield. Shields are boards that can be plugged on top of the Arduino PCB extending its capabilities. The ArduSiPM Shield (Fig. 1) is our custom designed board with all electronics interface from Arduino DUE and a SiPM photodetector. The ArduSiPM Shield plugged in the Arduino DUE give a compact system including Front end electronics and data acquisition system (Fig. 2).

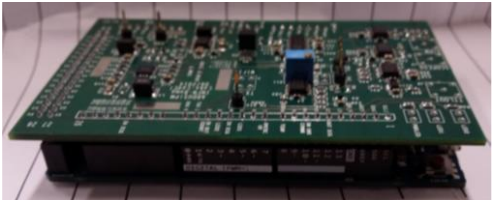

Fig. 2 The ArduSiPM (ArduinoDUE+ArduSiPM Shield)

The global architecture of the system is in Fig. 3,

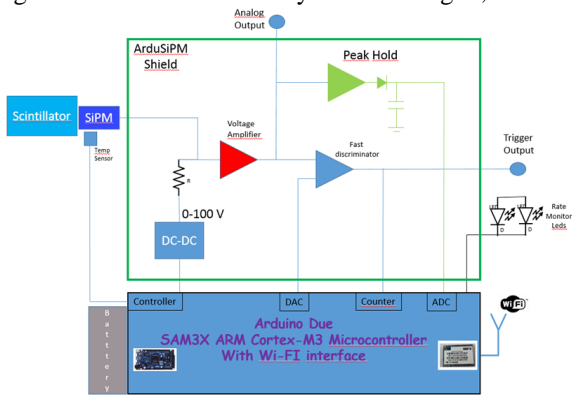

Fig. 3 ArduSiPM Block Diagram

there are an external SiPM with a temperature sensor, an internal digital controlled DC-DC converter as voltage supply of the SiPM , a voltage amplifier , a fast discriminator with a programmable threshold, a peak hold circuit for pulse amplitude, leds for monitoring, all outputs from analog circuit and digital controls are connected to the Arduino DUE board. The main components of the ArduSiPM Shield are:

### A. DC-DC converter.

The DC-DC converter use the main supply of 5 V to generate a supply in the range of 30-100 V, fixed the nominal voltage it is possible to vary around this value of few Volt with an 8 bits resolution. The temperature variation of the SiPM gain is one of the main problem of this kind of detector.

For each type of SiPM is tabulated the gain variation vs temperature variation. The amplification of the SiPM varies also in know function of the power supply voltage around the nominal value, using the fine regulation of the voltage in the ArduSiPM we can stabilize the gain of the detector using as input parameter the SiPM temperature read from the temperature sensor.

### B. Voltage Amplifier.

A Voltage preamplifier read the signal from the SiPM detector and adapt the value to the range of fast discriminator and to the range of Arduino DUE SAM3X8E Analog to Digital converter.
The Amplifier is fast to follow the SiPM response of few ns and linear in all the range of SiPM signal, with a noise less than a single SiPM single pixel signal to guarantee an accurate discrimination and amplitude measurement. A replicated output of the amplifier is present as external connection as Analog Output. It can be used with a scope to study the waveform in output from the SiPM.

### C. Fast Discriminator.

The Signal coming from the SiPM is very short (few ns) a fast 7 ns discriminator is used to discriminate over threshold signals and count them using Arduino DUE SAM3X8E counter. The Threshold value is digitally controlled and his value is monitored using one channel of SAM3X8E ADC. The width of discriminator is programmable to avoid after pulse counting and control the death time of the pulse acquisition. A replicated TTL output of the fast discriminator is present as Trigger Output. It can be used to trigger external acquisition system.

### D. Peak Hold circuit.

A precise circuit with fast peak detector is used as peak hold. A sampling comparator features a very short switching charge to measure the short pulses coming out from the SiPM detector when the pulse generate a trigger. The Pulses are stretched over 1 μs to be converted from the 1 MSPS 12 Bits ADC of the SAM3X8E. A programmable digital signal control the fast discharge circuit to reset the circuit of the peak hold after the ADC acquisition, and to rearm the system for a new acquisition

### E. Monitoring LEDs.

There are two LEDs directly controlled from Arduino DUE, in our software implementation one led flash every over threshold pulse and another every ten pulses in one second windows.

## III. ARDUINO DUE FUNCTION IN ARDUSiPM

The Arduino DUE main component is the SAM3X8E a System on Chip (SoC) [2] with an ARM Cortex-M3 CPU core and complex peripheral interfaces. The main peripheral for our design are:
- 16 Channel Multiplexed Analog to digital converter 12 bits 1 MSPS,
- Multiple input output pins
- 9 Timer used as Counter or pulse generator
- 2 Digital to Analog converter with 12 bits resolution
- Different serial interface like I2C, SPI, onewire, RS232

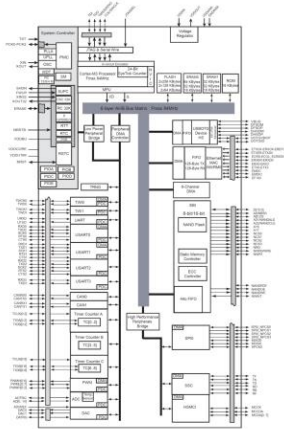

Fig. 4 SAM3X8E System on Chip Architecture[2].

The CPU and the peripheral are programmed using an easy to use development software, with high-level instruction, for main program and interrupt handling, with the possibility to use all the complex features of the SoC. The block used and the function used to control and acquire the ArduSiPM Shield are:

### A. The Timer counter module

There are 3 Timer module with three blocks each one, any block can be used in waveform mode to generate waveform of digital signals and in capture mode as 32 bits counter.

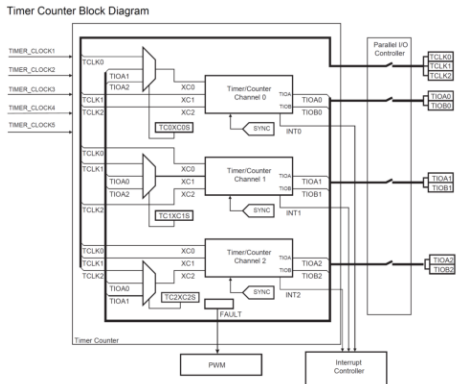

Fig. 5 Timer Counter Block Diagram [2]

These Timer Blocks are very flexible and can run after the programming as independent hardware without the direct control of the CPU core. We use this modules in waveform mode to generate the one second tick time, the reset signals for the discriminator and the peak hold circuit, the two signal to control the LED flash. We use the Timer in Counter mode for the rate measurement of the signal over threshold and for the 25 ns precision, Time to digital conversion used for absolute time measurement.

### B. The Analog to Digital converter (ADC) module

The SAM3X8E contains an Analog to Digital Converter block (Fig. 6) with a 12 bits resolution and with a maximum sample rate of 1 MSPS. The ADC has 12 multiplexed Channel, with a sample and hold and a programmable gain amplifier, two channel are read via software with a slow sample rate to monitor the bias voltage and the threshold voltage of the discriminator. One Channel is used at the maximum conversion rate to acquire the amplitude at the output of peak hold circuit. The ADC provides an interrupt for the main CPU to read out the value after the conversion.

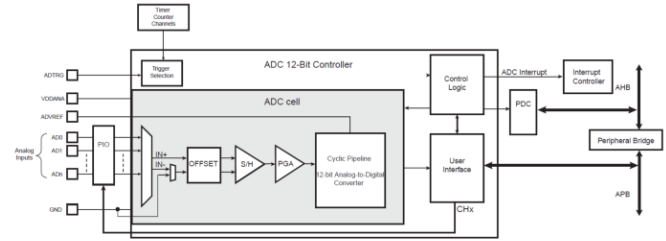

Fig. 6 SAM3X8e Analog to Digital Converter block [2].

### C. The Digital to Analog Conversion Module

The SAM3X8E contains as peripheral two Digital to Analog converter channel (Fig. 7). In the ArduSiPM one of this channel, control the threshold value of the discriminator. The other channel can be used for the coarse setting of the SiPM voltage supply.

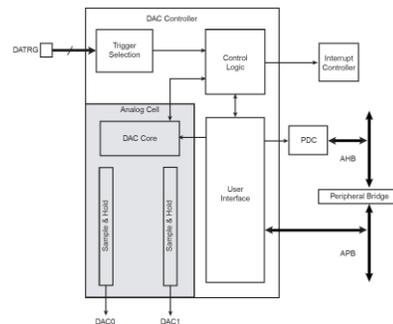

Fig. 7 SAM Digital to Analog Converter Block [2].

### D. Serial Communication.

There are different serial communication channel to communicate with devices. Te onewire port is used to read the temperature sensor, the i2c can be used for an external LCD display and the RS232 serial interface at the speed of 200 KBs is the port to export the mesured data and the monitoring parameter and to send all the configuration command. An internal Ethernet MAC is present but not pin available in the original Arduino Due design. There are Arduino DUE clone were this port is available and can be used to improve the communication speed at Mbs without changing the ArduSiPM

Shield design. With the ethernet port we can also implement the NTP (Network Time Protocol) to have a precise absolute time to correlate in time the readout of multiple ArduSiPM.

*E. Wi-Fi interface*

In this implementation, we use as interface with the external world a small factor 2x3 cm serial to Wi-FI module. The module has embedded a high performance 360 MHz MIPS24KEc CPU core, for TCP/IP serial communication. The module can be configured as an infrastructure point, in this case the ArduSiPM can be accessed as a point of a complex network (also available via internet) or in peer to peer mode for direct connection to the client.

*F. ArduSiPM Tablet and PC Client*

The communication protocol of the ArduSiPM use a light textual based interface and can be used like a terminal. This is a powerful for expert user but can be heavy for inexpert. The lack of Graphical user interface can be brightly resolved using as client a PC or a Tablet. We develop an Android App that can control and display data in raw or in graphical way after post processing. As example there is also an interface that a surgeon can use in an operation theater.

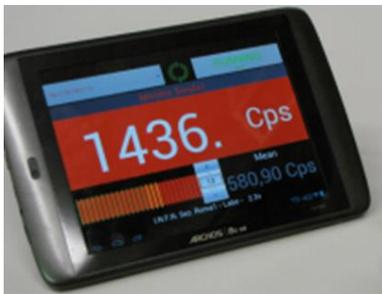

Fig. 8 Example of Android App client with rate display and sound alarms.

IV. ARDUSIPM DATA MEASURAMENT.

ArduSiPM is a data acquisition system for SiPM and provide measurments.

*A. ArduSiPM measurements output*

ArduSiPM split the continuous time in fixed acquisition windows. The default value of each window is one second in this case the value of number of pulses is coincident with the frequency value.
The pulse parameter: relative time and amplitude, are stored in continuous after the arrival of each pulse using interrupt routine. The print is done as soon as possible when the rs232 channel is free, the number of pulses arrived is printed at the end of each acquisition window. The buffer act as derandomizer to optimize the data output in the time window. The system is designed to running with negligible dead time. Any one of this measurement can be enabled or not

- The arrival time respect the start of the time window with a precision up to 25 ns.
- The amplitude of each pulse with a 12 bits resolution
- The number of pulse for each time window

The Fig. 9 shows the measurements in amplitude A1,A2,A3… and in arrival time t1,t2,t3,…..

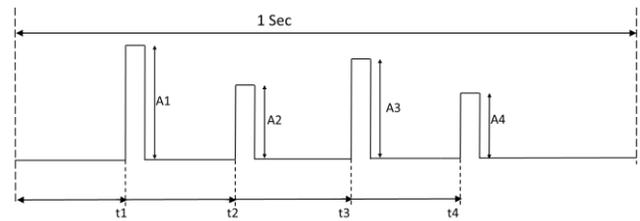

Fig. 9 Measurements in an acquisition window

The output format is readable (Fig. 10), to optimize the RS232 channel the amplitude and the arrival time are in hexadecimal in variable format with MSB zero suppression, the rate is printed in decimal format.The maximum data acquisition rate depends from the data enabled in output; it is of the order of:
- 20 MHz in rate mode only.
- 4-6 KHz with amplitude value
- 1-2 KHz with arrival time.

```
Only rate:
$10
$50
$244

ADC+Rate:
v1Fv1Dv22v27v1Dv19v20v23v20v1Cv19v1F$12
v18v1Ev1Ev1Bv19v1Bv29v19v1Av1Dv1Bv1Dv2Av18v1B$15
v15v20v21v21v1Dv1Fv1Av1Av1A$9
v19v17v1Bv18v1Cv1Dv1D$7

TDC+ADC+RATE:             Legend:
taedvataf0v7tv9v3$3       vXXX ADC Value in HEX MSB zero suppressed
                          tXXXXXXXX TDC value in HEX MSB zero suppressed
                          $XXX rate in Hz
```

Fig. 10 Example of data stream comings out from the ArduSiPM

*B. Real measurements.*

We test our system to understand the performance, The digital part gives a perfect correspondence in rate and time measurements by construction, the analog chain: voltage Amplifier, peak hold circuit and ADC show a very good linearity as we measure in Fig. 11.

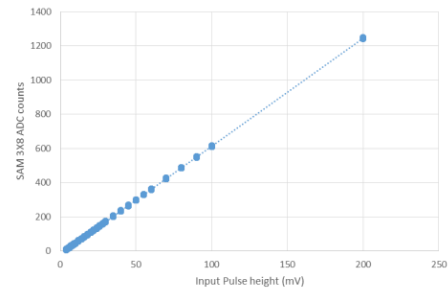

Fig. 11 test of analog chain linearity injecting a calibration pulse and reading the ADC value

Using a BC408 crystal coupled with a SiPM SensL MicroSL-10035-X18 with a pixel size of 35 μm and 576 pixel we acquire the spectrum of the Sr90 radioactive source in Fig. 12.

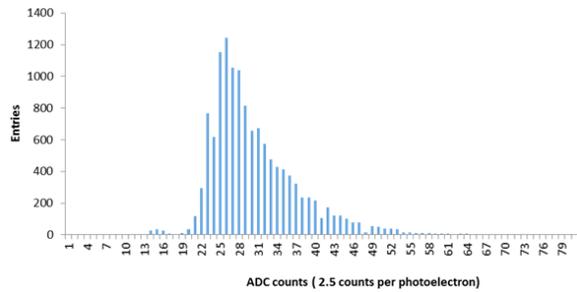

**Fig.** 12 Sr90 spectrum using a BC408 plastic scintillator and a SensL SiPM.

We test the system with SensL and Hamamtsu SiPM coupled with different Scintillator.

## V. EXAMPLE OF REAL APPLICATION.

The ArduSiPM is a flexible system to control and acquire SiPM. We use this system in real application in the field of Healthcare , Beam monitoring and in various testbench.

### A. Use of ArduSiPM as monitor system of an intraoperative beta probe for tumor discover.

The radio-guided surgery (RGS) is a technique that enables the surgeon to perform complete tumor resections, minimizing the amount of healthy tissue removed.

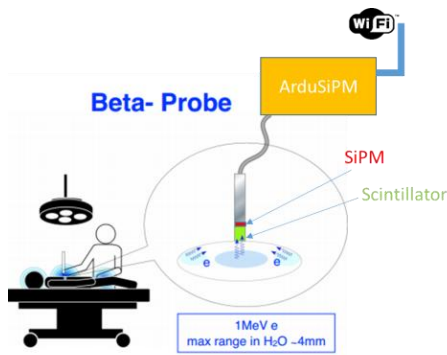

Fig. 13 ArduSiPM in a radioguided surgery beta proble.

We use the ArduSiPM electronics in a Radioguided surgery beta probe[3]. in this particular case we use as detector a para-Therphenil scintillator coupled with SiPM detector.
Para-Therphenil Light Yeld is 30 000 photons/MeV and is Trasparent to gamma due to low density. In this application a small piece of scintillator produces tens of photons when is crossed from a beta particle emitted from a Y90 labelled radiotracer.
The use of the SiPM detector and the ArduSiPM readout unit respect of the previous system with a Photo Multiplier awesome improve the cost and the portability of the system that can run for more than ten hours with a Lithium polymer battery

### B. Beam monitor and trigger unit of UA9 CERN experiment.

The UA9 collaboration[4] is investigating how tiny bent crystals could improve how beams are collimated in modern hadron colliders such as the LHC.

The planes in crystalline solids can constrain the directions that charged particles take as they pass through. Physicists can use this "channelling" property of crystals to steer particle beams. The old beam monitor of the UA9 at H8 CERN experiment was built using big scintillator with Photo Multiplier and NIM electronics and counters. We use a 10 x10 mm BC408 plastic scintillator coupled with a SensL SiPM to trigger together with the thin trigger [5]the arrival time of the beam (using the trigger output of the ArduSiPM) Fig. 14 and monitor the beam time profile(Fig. 15).

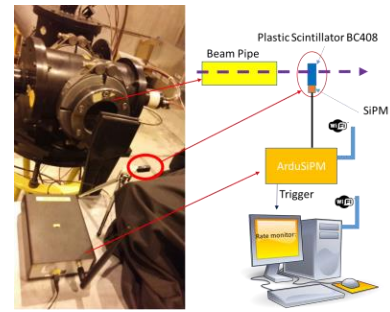

Fig. 14 Experiment setup of ArduSiPM in the UA9 experiment (H8 area).

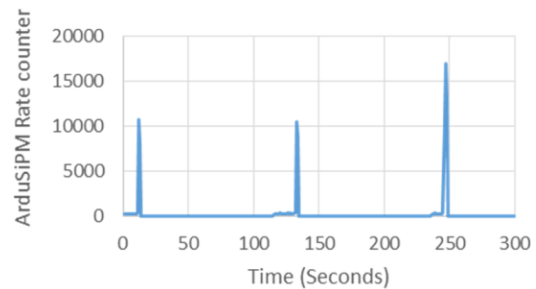

Fig. 15 Time profile of the UA9 beam acquired with ArduSiPM.

The small size of the scintillator and the coupled SiPM allows a direct contact with the UA9 apparatus for a precise spot identification of the incoming beam.

## ACKNOWLEDGMENT

We want to thanks Francesco Di Lorenzo for one of the first Schematic layout of the ArduSiPM, Riccardo Lunadei for the PCB layout, Riccardo Faccini, Silvio Morganti Elena Solfaroli, Andrea Russomando for the usefull discussion about the Android interface for the Beta probe, the UA9 collaboration for the support in the installation in their apparatus.

## REFERENCES

[1] http://arduino.cc/en/pmwiki.php?n=Main/ArduinoBoardDue
[2] AT91 SAM3x8e Atmel corporation data sheet.
[3] Solfaroli Camillocci, E. et al. A novel radioguided surgery technique exploiting β- decays. Sci. Rep. 4, 4401; DOI:10.1038/srep04401 (2014).
[4] Scandale, W et al. Observation of focusing of 400 GeV/c proton beam with the help of bent crystals", Pubblicazione su rivista: PHYSICS LETTERS B; Volume 733, 2 June 2014, Pages 366–372
[5] Iacoangeli, F., Bocci, V. ; Cavoto, G. ; Recchia, L. The Thin Light Trigger for the UA9 experiment, DOI:10.1109/NSSMIC.2013.6829750
[6] Bondarenko G et al 2000 Limited Geiger-mode microcell silicon photodiode: new results Nucl. Instrum. Methods Phys. Res. A 442 187–92.